\documentstyle[prl,multicol,aps,epsf]{revtex}
\begin{document}
\epsfverbosetrue
\title{Stability of vortex solitons in a photorefractive optical lattice}
\author{Jianke Yang}
\address{Department of Mathematics and Statistics, University of Vermont,
Burlington, VT 05401}
\maketitle

\begin{abstract}
Stability of off-site vortex solitons in a photorefractive optical lattice
is analyzed. It is shown that such solitons are linearly unstable in both the high and 
low intensity limits. In the high-intensity limit, the vortex looks
like a familiar ring vortex, and it suffers oscillatory instabilities. 
In the low-intensity limit, the vortex suffers both oscillatory and
Vakhitov-Kolokolov instabilities. However, in the moderate-intensity regime, 
the vortex becomes stable if the lattice intensity or the applied voltage
is above a certain threshold value. Stability regions of vortices
are also determined at typical experimental parameters. 

%190.0190: Nonlinear optics
%190.5330: Photorefractive nonlinear optics
\end{abstract}

\begin{multicols}{2}
\narrowtext
Vortex solitons are ubiquitous in many branches of physics such as optics 
\cite{kivshar} and
Bose-Einstein condensation \cite{ketterle}. In a homogeneous medium, bright vortex
rings are unstable \cite{firth}, and only dark vortex solitons are possible with defocusing
nonlinearity \cite{kivshar,swartzlander}. 
However, in the presence of a periodic optical lattice, stable lattice vortices
become possible due to the guiding properties of the lattice. 
Indeed, recent theoretical work \cite{Malomed,YangZiad} has shown that 
in an optical lattice with Kerr nonlinearity, 
both on-site vortices (vortices whose singularity is located on a lattice site)
\cite{Malomed} and off-site vortices 
(vortices whose singularity is located between sites) \cite{YangZiad} 
are stable within certain ranges of parameters. 
These theoretical studies are quickly followed by experiments in
photorefractive crystals, where vortex lattice solitons were observed very recently
\cite{ChenKivshar,Segev}. 
For a review of other nonlinear localized states in one- and two-dimensional 
periodic optical waveguides, see \cite{ref1,ref2}. 

Stability of vortex lattice solitons in photorefractive crystals
is clearly an important issue. This question was considered
in \cite{Segev}, where the evolution of a particular on-site lattice vortex
under random-noise perturbations was simulated. 
It was found that the on-site vortex was stable to very long distances. 
However, we know that lattice vortices in photorefractive crystals
can not be all stable. For instance, when the peak intensity (or power)
of the vortex is high, the lattice is effectively weak, thus
the lattice vortex would become the familiar ring vortex,
which is known to be unstable [see Fig. 1(b) below] \cite{firth}. 
The natural questions to ask then are: what 
lattice vortices are stable? If lattice vortices are unstable, 
what are the sources of their instability?
So far, these questions have not been addressed comprehensively 
for either of the on-site and off-site lattice vortices. 

In this paper, we study the off-site vortex solitons in a photorefractive optical lattice. 
Off-site lattice vortices are more closely packed ---
the diagonal distance between their four main lobes is $\sqrt{2}$ times
shorter than that of on-site vortices. Thus their 
dynamics is stronger and more interesting. 
We show that these vortices are not only unstable in the 
high-intensity limit, but also in the low-intensity limit. 
However, they do become stable in the moderate-intensity regime
if the lattice intensity or the applied voltage reaches over a certain threshold. 
We also determine the stability regions of vortices
for a wide range of experimental parameters, and show that
the stability region expands when the applied voltage increases. 

The mathematical model for light propagation in a photorefractive crystal
has been known for some time \cite{efremidis}. 
Here we make the usual paraxial assumption, and the assumption that the photorefractive
screening nonlinearity acts isotropically along the two transverse directions, 
both of which are justified in many experiments. 
If the probe beam is extra-ordinarily polarized, while
the lattice is ordinarily polarized, then the probe beam does not affect 
the linear lattice. 
In this case, the governing equation for the probe beam is \cite{efremidis}
\begin{equation} \label{lattice_model}
iU_z+\frac{1}{2k_1} \left(U_{xx}+U_{yy}\right)
-\frac{1}{2}k_0n_e^3r_{33}E_{sc}U=0, 
\end{equation}
where $U$ is the slowly-varying amplitude of the probe beam, 
$z$ is the distance along the direction of the crystal, 
$(x, y)$ are distances along the transverse directions, 
$k_0=\frac{2\pi}{\lambda_0}$ is the wavenumber of the laser in the vacuum
($\lambda_0$ is the wavelength), 
$n_e$ is the refractive index along the extraordinary axis, $k_1=k_0n_e$, 
$r_{33}$ is the electro-optic coefficient for the extraordinary polarization, 
$E_{sc}$ is the space-charge field, 
%\begin{equation} \label{Esc}
$E_{sc}=E_0/[1+I_l(x,y)+|U|^2]$,
%\end{equation}
$E_0$ is the applied DC field, and 
$I_{l}$ is the field intensity of the optical lattice. 
Here the intensities of the probe beam and the lattice have been normalized
with respect to the dark irradiance of the crystal $I_d$. 
Material damping of the probe beam is very weak in typical experiments 
since the crystals are fairly short (up to 2 cm).  Hence it is neglected in 
Eq. (\ref{lattice_model}). 
If the lattice is periodic along the $x$ and $y$ directions (rectangular lattice), 
%then $I_{l}$ can be expressed as
then
%\begin{equation} \label{lattice_intensity}
%I_{l}(x,y)=I_0\sin^2\frac{\pi}{D}x\sin^2\frac{\pi}{D}y, 
$I_{l}(x,y)=I_0\sin^2\left(\pi x/D\right) \sin^2\left(\pi y/D\right),$
%\end{equation} 
where $I_0$ is its peak intensity, and $D$ is its spacing.   

Eq. (\ref{lattice_model}) can be non-dimensionalized. If we
measure the transverse directions $(x,y)$ in units of
$D/\pi$, the $z$ direction in units of $2k_1D^2/\pi^2$,
and the applied voltage $E_0$ in units of
$\pi^2/(k_0^2n_e^4D^2r_{33})$, then 
Eq. (\ref{lattice_model}) becomes
\begin{equation} \label{model}
iU_z+U_{xx}+U_{yy}-\frac{E_0}{1+I_0\sin^2x\sin^2y+|U|^2}U=0.
\end{equation}
Consistent with the experiments \cite{ref2}, 
we choose physical parameters as 
$D=20\mu m, \lambda_0=0.5\mu m, n_e=2.3, r_{33}=280 pm/V.$
Thus, in this paper, one $x$ or $y$ unit corresponds to $6.4 \mu m$, 
one $z$ unit corresponds to 2.3 mm, and one $E_0$ unit corresponds
to 20 V/mm in physical units. 

Lattice vortices of Eq. (\ref{model}) are sought of in the
form $U=u(x, y) e^{-\mu z}$, where $\mu$ is the propagation constant. 
%Then $u(x, y)$ satisfies the nonlinear equation
%\begin{equation} \label{u}
%u_{xx}+u_{yy}+\left(\mu-\frac{E_0}{1+I_0\sin^2x\sin^2y+|u|^2}\right)u=0.
%\end{equation}
We determined these vortices by a Fourier iteration method \cite{YangZiad}. 
At lattice intensity $I_0=I_d$ and applied voltage $E_0=8$, these vortices
are shown in Fig. 1. 
We see that when the vortex' peak intensity $I_p$ is high, 
the vortex becomes a familiar ring vortex [see Fig. 1(b)] since
the optical lattice is relatively negligible in this case. 
As $I_p$ decreases, the vortex develops four
major lobes at four adjacent lattice sites 
in a square configuration [see Fig. 1(c,e)]. When $I_p$ 
is low, the vortex spreads over to more lattice sites and becomes
less localized [see Fig. 1(f)]. The phase fields of all these
lattice vortices, however, remain
qualitatively the same as in a regular ring vortex [see Fig. 1(d)]. 
Note that the singularities (centers) of these vortices
are not on a lattice site, thus these vortices are off-site vortices. 
An interesting fact we found is that, for given lattice intensity 
and applied voltage values, lattice vortices with
$I_p$ below a certain threshold $I_{p,c}$ do not exist. 
In the present case where $I_0=I_d$ and $E_0=8$, this threshold value is
$I_{p,c}\approx 0.28I_d$. This fact indicates that, 
unlike fundamental lattice solitons, 
lattice vortices do not bifurcate from infinitesimal Block waves. 

%As we have seen in Fig. 1, in the high-intensity limit, vortices
%in photorefractive optical lattices approach the 
%lattice-free ring vortices. This is different from vortices
%in a Kerr medium, where they approach four singular spikes
%in the high-intensity limit \cite{YangZiad}. 
%The reason is that, in a Kerr medium, fundamental solitons become
%narrower and narrower when their peak intensities get higher and higher. 
%But in a photorefractive crystal where the nonlinearity is saturable, 
%fundamental solitons flatten out when their intensities become high
%(see Fig. 1a of \cite{YangPRE02}). Thus in a photorefractive
%lattice, the four lobes of the vortex join together and form 
%a ring vortex at high intensities, while in a Kerr lattice, 
%the four lobes develop into four singular spikes at high intensities. 

We can further determine the power and peak-intensity diagrams
versus the propagation constant $\mu$. Here the power is defined 
as $P\equiv \int_{-\infty}^\infty \int_{-\infty}^\infty |u|^2 dxdy$. 
When $I_0=I_d$ and $E_0=8$, 
the results are shown in Fig. 2(a). We see that
the peak-intensity is a monotone-decreasing function of $\mu$, 
but the power is monotone-decreasing only when the peak intensity
is above $0.34I_d$. Below this intensity value, the power starts to 
increase with $\mu$. This behavior qualitatively holds also for other 
$I_0$ and $E_0$ values, and it is similar to that 
in the Kerr medium \cite{YangZiad}. 

Now we address the critical question of linear stability of
these vortices in a photorefractive lattice. 
High-intensity lattice vortices clearly should be linearly unstable because
they approach the regular ring vortex [see Fig. 1(b)] \cite{firth}. 
The instability is oscillatory (i.e., the unstable eigenvalues are complex). 
At low intensities, $dP/d\mu>0$, hence the lattice vortices
are expected to be linearly unstable as well according to the
Vakhitov-Kolokolov (VK) criterion \cite{VK}. The
VK instability is purely exponential (i.e., the unstable eigenvalues are
purely real). How about the stability behaviors of vortices at 
moderate peak intensities? To answer this question, 
we have simulated the linearized equation of (\ref{model}) around
lattice vortices $u(x, y) e^{-\mu z}$ for very long distances, 
and obtained the unstable eigenvalues $\sigma$ of small disturbances
(the real part of $\sigma$ is the growth rate). 
The results for $I_0=I_d$ and $E_0=8$ are shown in Fig. 2(b). We find that
lattice vortices are linearly unstable when 
$I_p>2.1I_d$ and $I_p<0.70 I_d$, consistent with our 
expectations. In addition, the instability for $I_p > 2.1I_d$ is oscillatory
[Im$(\sigma) \ne 0$], and the VK instability for $I_p < 0.34I_d$ is purely 
exponential [Im$(\sigma)=0$], as we would expect. 
However, Fig. 2(b) reveals another oscillatory instability
for $0.34I_d<I_p<0.70I_d$, which was not anticipated. 
This additional oscillatory instability has been seen
in the Kerr medium before \cite{YangZiad}. 

A more important result revealed by Fig. 2(b) is that 
for $0.70 I_d <I_p <2.1I_d$, 
lattice vortices are linearly stable. This is an important result. 
It implies that lattice vortices with such moderate intensities 
could be observable in experiments. 

When lattice vortices are linearly unstable, what is the outcome of the
instability? To address this question, 
we select the linearly-unstable
vortex soliton with $I_0=1, E_0=8$ and peak intensity $I_p=3I_d$, and perturb
it by random noise. The noise has Gaussian intensity distribution in the 
spectral $k$-space with FWHM 2 times larger than the soliton FWHM spectrum. 
The noise power is 1\% of the soliton's. The simulation result on the
evolution of this vortex under noise perturbations is shown in Fig. 3. 
We see that this vortex breaks up into a fundamental lattice soliton plus
some radiation. This breakup scenario is typical of unstable 
lattice vortices under noise perturbations.  

When lattice vortices are linearly stable, how do they evolve nonlinearly?
To answer this question, we select the linearly-stable
vortex soliton with $I_0=1, E_0=8$ and $I_p=1.5I_d$, 
and perturb it by the same random noise as described above. 
The simulation result on the evolution of this perturbed vortex is 
shown in Fig. 4. We see that this vortex does propagate stably. 
In addition, its phase structure is maintained throughout the evolution. 
Evolution of other linearly-stable lattice vortices under weak 
perturbations is similar. This means that linearly-stable vortex solitons
could be observed in experiments, as the work \cite{ChenKivshar,Segev}
has shown. 

Above at specific lattice intensity and applied voltage values $I_0=I_d$ and $E_0=8$,
we have revealed the sources of instability of lattice vortices, and 
obtained stable lattice vortices. The next question quickly follows: 
if the lattice intensity and voltage values are varied, how would they affect the
stability properties of lattice vortices? 
%or equivalently, 
%at general lattice intensity and voltage values, what lattice vortices are 
%linearly stable? 
This question is important for experiments. 
To find the answer to this question, we have systematically determined
the linear stabilities of lattice vortices at various lattice intensity,
applied voltage and vortex peak-intensity values. The results are summarized
in Fig. 5. Here at two applied voltage values $E_0=8$ and 10, the stability boundaries
are presented in the $(I_p, I_0)$ plane. 
This figure reveals several important facts. 
First, high-intensity and low-intensity vortex solitons are always
linearly unstable, as we have observed in Fig. 2 above. 
Second, when the applied-voltage value $E_0$ is fixed,  
there is a threshold lattice intensity $I_{0,c}$, below which
all lattice vortices (including moderate-intensity ones)
are linearly unstable. When $E_0=8$, this threshold value is
$I_{0,c}\approx 0.7I_d$; while when $E_0$ is increased to 10, 
$I_{0,c}$ decreases to $0.44I_d$.
Similarly, when the lattice intensity $I_0$ is fixed, 
there is also a threshold applied-voltage value below which
all lattice vortices are linearly unstable. 
Thirdly, when the applied voltage increases, the region of
stable lattice vortices expands. In other words, 
higher-applied voltage stabilizes lattice vortices. 
Fig. 5 should be helpful to experimentalists
on their choices of physical parameters for the observation
of lattice vortices. 

%In this paper, the stability analysis was presented for off-site vortex
%solitons in photorefractive optical lattices. 
%A similar analysis can be carried out for on-site vortices as well, and
%this will be pursued in the future. 

In summary, we have carried out a stability analysis
on off-site lattice vortices in photorefractive optical lattices. 
We showed that high- and low-intensity lattice vortices suffer
oscillatory and VK instabilities, but moderate-intensity 
vortices can be stable when the applied voltage or lattice intensity
is above a certain threshold. Higher applied voltage stabilizes
lattice vortices.

\vspace{-0.3cm}
\begin{figure}
\begin{center}
\setlength{\epsfxsize}{6cm} \epsfbox{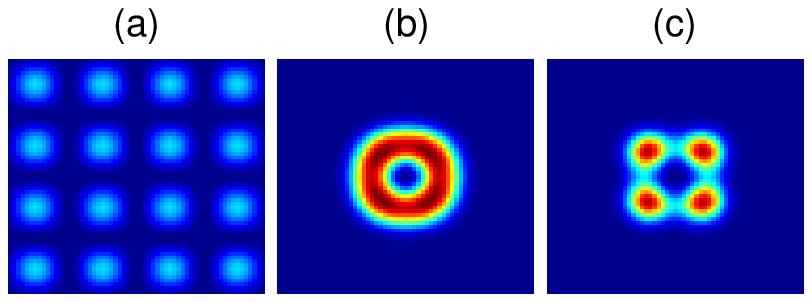}

\setlength{\epsfxsize}{6cm} \epsfbox{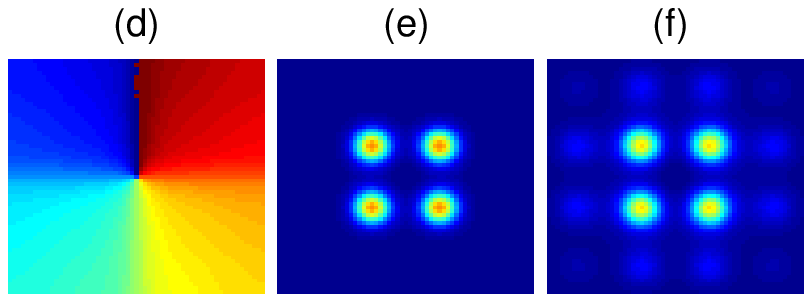}

\vspace{0.2cm}
\caption{(a) Intensity field of the optical lattice with $I_0=I_d$; 
(b, c, e, f) intensity fields of lattice vortices with 
peak intensities 15, 5, 1.5 and 0.3 $I_d$ respectively under the applied voltage $E_0=8$; 
(d) phase structure of these vortices. }
\end{center}
\end{figure}

\vspace{-0.3cm}
\begin{figure}
\begin{center}
\setlength{\epsfxsize}{3.7cm} \epsfbox{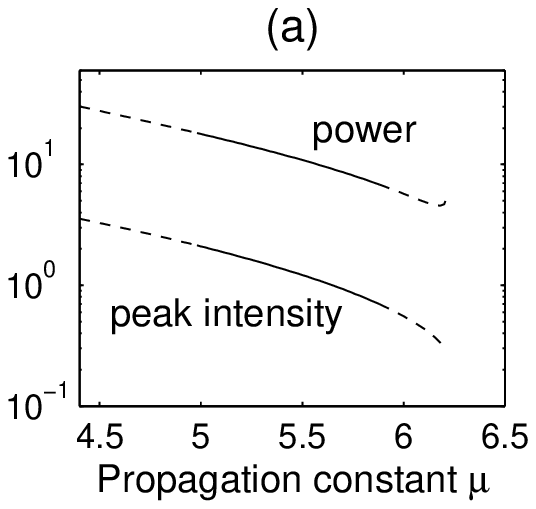} \hspace{0.2cm}
\setlength{\epsfxsize}{3.9cm} \epsfbox{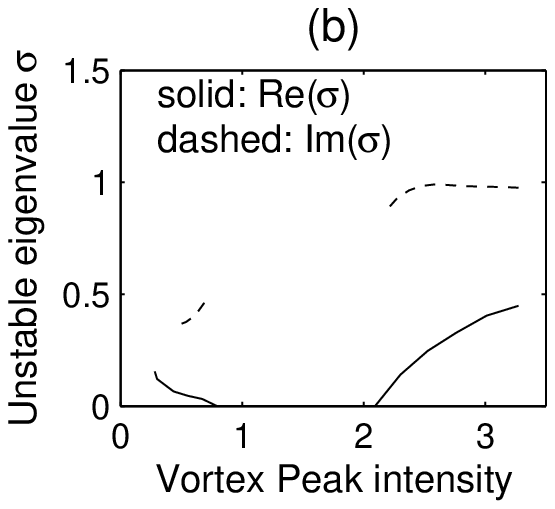}

\vspace{0.2cm}
\caption{(a) Power and peak-intensity diagrams of lattice vortices
at $I_0=I_d$ and $E_0=8$; solid-line portion: stable vortices; 
dashed-line portions: unstable vortices; 
(b) unstable eigenvalues of these vortices versus their peak intensity. }
\end{center}
\end{figure}

\vspace{-0.3cm}
\begin{figure}
\begin{center}
\setlength{\epsfxsize}{8cm} \epsfbox{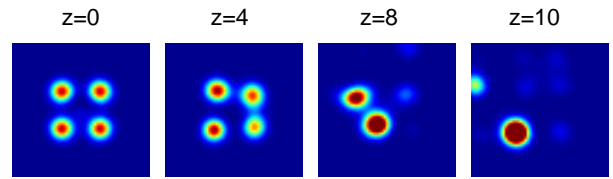}

\vspace{0.2cm}
\caption{Break-up of a lattice vortex soliton
with $I_0=I_d$, $E_0=8$ and $I_p=3I_d$ under random-noise perturbations. }
\end{center}
\end{figure}

\vspace{-0.3cm}
\begin{figure}
\begin{center}
\setlength{\epsfxsize}{6cm} \epsfbox{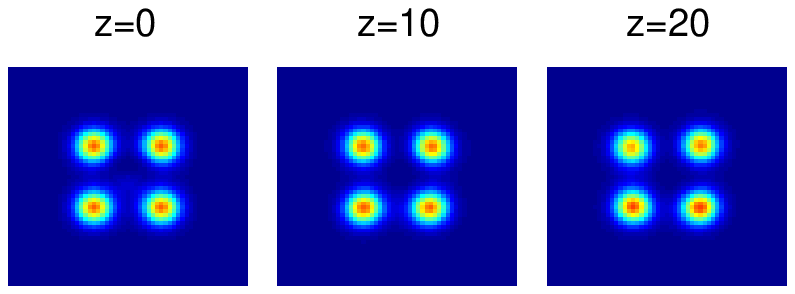}

\setlength{\epsfxsize}{6cm} \epsfbox{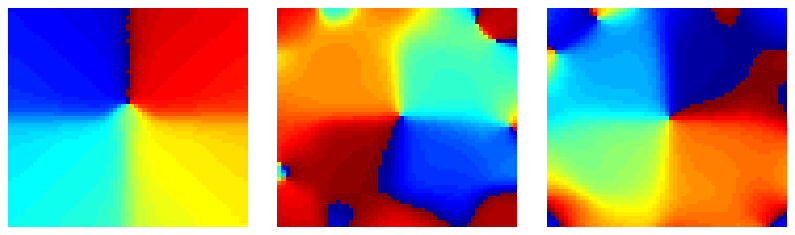}

\vspace{0.2cm}
\caption{Stable propagation of a lattice vortex soliton
with $I_0=I_d$, $E_0=8$ and $I_p=1.5I_d$ under random-noise perturbations. 
Top row: intensity; bottom row: phase.}
\end{center}
\end{figure}

\begin{figure}
\begin{center}
\setlength{\epsfxsize}{4.2cm} \epsfbox{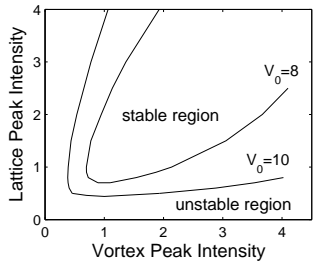}

\vspace{0.2cm}
\caption{Stability boundaries of lattice vortex solitons in the
$(I_p, I_0)$ plane at two applied voltages $E_0=8$ and 10. }
\end{center}
\end{figure}

\end{multicols}

\end{document}